\documentclass[journal,comsoc]{IEEEtran}

\ifCLASSINFOpdf
\else
\fi
\usepackage[cmex10]{amsmath}
\usepackage{amssymb}
\usepackage{siunitx}
\usepackage{caption}
\usepackage{mathtools}
\DeclarePairedDelimiter{\ceil}{\lceil}{\rceil}
\usepackage{multirow}
\usepackage{algpseudocode}
\usepackage{algorithm}
\algnewcommand{\Initialization}[1]{%
  \State \textbf{initialization:}
  \Statex \hspace*{\algorithmicindent}\parbox[t]{.8\linewidth}{\raggedright #1}
}

\algnewcommand{\Rep}[1]{%
  \State \textbf{repeat:}
  \Statex \hspace*{\algorithmicindent}\parbox[t]{.8\linewidth}{\raggedright #1}
}

\usepackage{tikz}
\usetikzlibrary{shapes,arrows,fit,positioning,calc}
\usetikzlibrary{plotmarks}
\usetikzlibrary{decorations.pathreplacing}
\usetikzlibrary{patterns}
\usetikzlibrary{arrows,automata}

\usepackage{pgfplots}
\pgfplotsset{compat=newest}

\begin{document}
%
\title{Continuous Phase Modulation With Faster-than-Nyquist Signaling for Channels With 1-bit Quantization and Oversampling at the Receiver}

\author{Rodrigo~R.~M.~de~Alencar,~\IEEEmembership{Student Member,~IEEE,} and Lukas~T.~N.~Landau,~\IEEEmembership{Member,~IEEE,}\vspace{-1em}		

\thanks{The authors are with Centro de Estudos em Telecomunica\c{c}\~{o}es Pontif\'{i}cia Universidade Cat\'{o}lica do Rio de Janeiro, Rio de Janeiro CEP 22453-900, Brazil, (email: \{alencar, lukas.landau\}@cetuc.puc-rio.br).} 
\thanks{This work has been supported by the {ELIOT ANR18-CE40-0030 and FAPESP 2018/12579-7} project.}
}

\maketitle


\begin{abstract}
Continuous phase modulation (CPM) with 1-bit quantization at the receiver is promising in terms of energy and spectral efficiency. In this study, CPM waveforms with symbol durations significantly shorter than the inverse of the signal bandwidth are proposed, termed faster-than-Nyquist CPM. 
This configuration provides a better steering of zero-crossings as compared to conventional CPM. Numerical results confirm a superior performance in terms of BER in comparison with state-of-the-art methods, while having the same spectral efficiency and a lower oversampling factor. Moreover, the new waveform can be detected with low-complexity, which yields almost the same performance as using the BCJR algorithm.
\end{abstract}



%

\vspace{-1em}
\section{Introduction}
Continuous phase modulation (CPM) yields spectral efficiency, smooth phase transitions and a constant envelope \cite{Anderson_1986,Sundberg_1986}, which allows for the use of energy efficient power amplifiers with low dynamic range.
At the receiver side, the energy consumption of the analog-to-digital converter (ADC) scales exponentially with the resolution in amplitude \cite{Walden_1999}.
Hence, in this study a low resolution ADC is considered, where the ADC provides only sign information about the received signal. 
In order to compensate for the loss in terms of the achievable rate, oversampling with respect to (w.r.t.) the signal bandwidth is considered.
In this context, it is shown that oversampling yields a significant gain in terms of achievable rate for the noiseless \cite{Shamai_1994} and for the noisy channel \cite{Landau_CL2017}. 

 As the information is implicitly conveyed in phase transitions of CPM signals, resolution in time is more promising than resolution of amplitude. 
 CPM signals with channels with 1-bit quantization and oversampling has been considered before in \cite{Landau_CPM_2018}, where the achievable rate is studied and maximized via optimization of sequences.
 Later, more practical approaches were proposed in \cite{Bender_SPAWC2019}, where the intermediate frequency and the waveform is considered in a geometrical analysis of the phase transitions.
 Moreover, in \cite{Alencar_2019}
it is presented how to exploit the channel with 1-bit quantization and oversampling by using iterative detection with sophisticated channel coding for CPM signals.  
 


In the present study, a CPM waveform is introduced
with a symbol duration that is only a fraction of the symbol duration of an equivalent CPFSK, which is promising in terms of construction of zero-crossings.
The proposed CPM waveform conveys the same information per time interval as the common CPFSK while its bandwidth can be the same and even lower.
Referring to the high signaling rate, like it is typical for 
\textit{faster-than-Nyquist} signaling \cite{Mazo_1975}, the novel waveform is termed faster-than-Nyquist continuous phase modulation (FTN-CPM).
Numerical results confirm that the proposed waveform yields a significantly reduced bit error rate (BER) as compared to the existing methods \cite{Bender_SPAWC2019, Landau_CPM_2018} with at least the same spectral efficiency. In addition, FTN-CPM can be detected with low-complexity and with a lower effective oversampling factor in comparison with the state-of-the-art methods.



The sequel is organized as follows: Section~\ref{sec:system_model} defines the system model, whereas Section~\ref{sec:proposed_wf} describes the proposed waveform. 
Section~\ref{sec:demodulation}
details the detection, which includes also the proposed low-complexity method.
Section~\ref{sec:numerical_results} discusses numerical results, while Section VI shows the conclusions.

 Sequences are denoted with $x^n= [x_1,\ldots,x_n]^T$. Likewise, sequences of vectors are written as $\boldsymbol{y}^n= [\boldsymbol{y}_1^T,\ldots,\boldsymbol{y}_n^T]^T$. A segment of a sequence is given by $x^{k}_{k-L}=[ x_{k-L}, \ldots,  x_k   ]^T$ and $\boldsymbol{y}^{k}_{k-L}=[ \boldsymbol{y}_{k-L}^T, \ldots,  \boldsymbol{y}_k^T   ]^T$.
\section{System Model}
\label{sec:system_model}
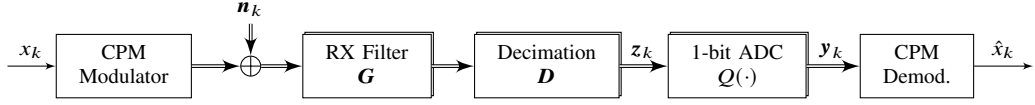
\begin{figure*}[h]
\begin{center}
\captionsetup{justification=centering}
\usetikzlibrary{shapes,arrows,fit,positioning,shadows,calc}
\usetikzlibrary{plotmarks}
\usetikzlibrary{decorations.pathreplacing}
\usetikzlibrary{patterns}

\tikzstyle{block} = [draw, fill=white, rectangle,minimum height=3em, minimum width=5.2em]	
\tikzstyle{block_rot} = [draw, fill=white, rectangle,minimum height=3.8em, minimum width=2em]
\tikzstyle{sum} = [draw, fill=white, circle, node distance=1em,path picture={\draw[black](path picture bounding box.south) -- (path picture bounding box.north) (path picture bounding box.west) -- (path picture bounding box.east);}]
\tikzstyle{coord} = [coordinate]

\tikzstyle{state}=[shape=circle,draw=blueTUD50,fill=blueTUD10]	
\tikzstyle{lightedge}=[<-,dotted]
\tikzstyle{mainstate}=[state,thick]
\tikzstyle{mainedge}=[<-,thick]

\tikzstyle{symbol}=[shape=circle,draw=blueTUD50,fill=blueTUD10,minimum width=1em,scale=0.6]	
\tikzstyle{sample1}=[shape=circle,draw,scale=0.3]	
\tikzstyle{sample2}=[shape=circle,draw,densely dashed,scale=0.3]	
\tikzstyle{interleave}=[shape=circle,draw,fill,minimum width=1em,scale=0.6]	

\tikzstyle{register} = [draw, fill=white, rectangle,minimum height=3em, minimum width=3em]	
\tikzstyle{mod2} = [draw, fill=white, circle, label={mod 2}, node distance=1em,path picture={\draw[black](path picture bounding box.south) -- (path picture bounding box.north) (path picture bounding box.west) -- (path picture bounding box.east);}]

\tikzset{multiple/.style={copy shadow={shadow xshift=0.1em,shadow yshift=0.1em}, draw=black,fill=white, rectangle,minimum height=3em, minimum width=5.2em},
multiple_rot/.style={copy shadow={shadow xshift=0.1em,shadow yshift=0.1em}, draw=black, fill=white, rectangle,minimum height=3.8em, minimum width=2em}}

\tikzset{>=latex,every picture/.style={font issue=\footnotesize},
         font issue/.style={execute at begin picture={#1\selectfont}}
        }

\begin{tikzpicture}[auto, node distance=4em,>=latex']
    
    \node[coord,  node distance=0em, align=center](source){};
	
    \node [block, right of=source,node distance=5.5em] (mod) {\begin{tabular}{c} CPM  \\  Modulator \end{tabular}};

    \node [sum,right of=mod,node distance=6em] (Sum1) {};
    \node [multiple, right of=Sum1,node distance=5.5em] (gt) {\begin{tabular}{c} RX Filter \\  $\boldsymbol{G}$ \end{tabular}};
		
    \node [multiple, right of=gt,node distance=8.5em] (decimation) {\begin{tabular}{c}Decimation\\  $\boldsymbol{D}$ \end{tabular}};
		
    \node [coord,label={$\boldsymbol{n}_k$},above of=Sum1,node distance=2em] (noise) {};
    \node [multiple, right of=decimation, node distance=9em] (adc) {\begin{tabular}{c} 1-bit ADC \\ $Q(\cdot)$ \end{tabular}};
    
    \node [block, right of=adc,node distance=8.5em] (demod) {\begin{tabular}{c} CPM  \\  Demod. \end{tabular}};
    
    \node [coord,node distance=5.5em, right of=demod] (output) {};
     
    \draw [->] (source) -- node {$x_k$} (mod);
    \draw [->,double] (mod) -- node {}(Sum1);
    \draw [->,double] (noise) -- node {}(Sum1);
    \draw [->,double] (Sum1) -- node {}(gt);
    \draw [->,double] (gt) -- node {}(decimation);	

    \draw [->,double] (decimation) -- node {$\boldsymbol{z}_k$}(adc);
		
    \draw [->,double] (adc) -- node {$\boldsymbol{y}_k$}(demod);
    \draw [->] (demod) -- node {$\hat{x}_k$} (output);
\end{tikzpicture}
\caption{ Discrete time description of the CPM system with 1-bit quantization and oversampling at the receiver}
\label{fig:system_model}       
\vspace{-1em}
\end{center}
\end{figure*}
The considered system model is based on the discrete time system model described before in \cite{Landau_CPM_2018}. Later in Section~\ref{sec:demodulation} the model is 
extended by different CPM demodulators for processing the quantized received signal, as illustrated in Fig.~\ref{fig:system_model}. In the sequel the individual building blocks are detailed.

\subsubsection{CPM Modulator}

The information conveying
phase term of the constant envelope CPM signal \cite{Anderson_1986}
reads
\begin{align}
\phi\left(t\right) = 2 \pi h  \sum_{k=0}^{\infty}  \alpha_{k}   f(t-k T_{\textrm{s}})  +\varphi_0   \text{,} 
\label{eq:cpm:phase_term}
\end{align}
where $T_{\textrm{s}}$ denotes the symbol duration, $h=\frac{K_{\textrm{cpm}}}{P_{\textrm{cpm}}}$ is the modulation index, $f\left(\cdot\right)$ is the phase response, $\varphi_0$ is a phase-offset and $\alpha_k$ represents the $k^{\textrm{th}}$ transmit symbol. For an even modulation order $M_{\textrm{cpm}}$, such transmit symbols are taken from an alphabet described by $\alpha_k \in \left\{  \pm 1, \pm 3,\ldots  ,\pm   (M_{\textrm{cpm}}-1)   \right\}$.
In order to obtain a finite number of phase states $K_{\textrm{cpm}}$ and $P_{\textrm{cpm}}$ must be relative prime positive integers.
The phase response function $f\left(\cdot\right)$ shapes the 
sequence of CPM symbols to the continuous phase signal with smooth transitions. The phase response is characterized by  
\begin{align}
f(\tau)=\begin{cases}
  0,  & \text{ if } \tau\leq 0{ ,}  \notag  \\  
  \frac{1}{2}, & \text{ if }   \tau >  T_{\textrm{cpm}}  \text{,}
\end{cases}   
\end{align}
where $T_{\textrm{cpm}}$ defines the CPM memory in terms of $L_{\textrm{cpm}}= \ceil{T_{\textrm{cpm}} / T_{\textrm{s}}}$ transmit symbols.
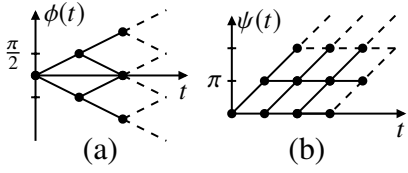
\begin{figure}[t]
\begin{center}
\captionsetup{justification=centering}
\tikzset{every picture/.style={line width=0.75pt}} 

\begin{tikzpicture}[x=0.75pt,y=0.75pt,yscale=-0.55,xscale=0.55]

\draw    (38,136) -- (38,18) (34,96) -- (42,96)(34,76) -- (42,76)(34,56) -- (42,56) ;
\draw [shift={(38,16)}, rotate = 450] [fill={rgb, 255:red, 0; green, 0; blue, 0 }  ][line width=0.75]  [draw opacity=0] (8.93,-4.29) -- (0,0) -- (8.93,4.29) -- cycle    ;

\draw    (38,76) -- (176,76) ;
\draw [shift={(178,76)}, rotate = 180] [fill={rgb, 255:red, 0; green, 0; blue, 0 }  ][line width=0.75]  [draw opacity=0] (8.93,-4.29) -- (0,0) -- (8.93,4.29) -- cycle    ;

\draw [line width=0.75]    (38,76) -- (118,36) ;
\draw [shift={(118,36)}, rotate = 333.43] [color={rgb, 255:red, 0; green, 0; blue, 0 }  ][fill={rgb, 255:red, 0; green, 0; blue, 0 }  ][line width=0.75]      (0, 0) circle [x radius= 3.35, y radius= 3.35]   ;
\draw [shift={(38,76)}, rotate = 333.43] [color={rgb, 255:red, 0; green, 0; blue, 0 }  ][fill={rgb, 255:red, 0; green, 0; blue, 0 }  ][line width=0.75]      (0, 0) circle [x radius= 3.35, y radius= 3.35]   ;
\draw [line width=0.75]    (78,56) -- (118,76) ;
\draw [shift={(118,76)}, rotate = 26.57] [color={rgb, 255:red, 0; green, 0; blue, 0 }  ][fill={rgb, 255:red, 0; green, 0; blue, 0 }  ][line width=0.75]      (0, 0) circle [x radius= 3.35, y radius= 3.35]   ;
\draw [shift={(78,56)}, rotate = 26.57] [color={rgb, 255:red, 0; green, 0; blue, 0 }  ][fill={rgb, 255:red, 0; green, 0; blue, 0 }  ][line width=0.75]      (0, 0) circle [x radius= 3.35, y radius= 3.35]   ;
\draw [line width=0.75]  [dash pattern={on 3pt off 3pt}]  (118,36) -- (158,56) ;

\draw [line width=0.75]    (78,96) -- (118,76) ;
\draw [shift={(118,76)}, rotate = 333.43] [color={rgb, 255:red, 0; green, 0; blue, 0 }  ][fill={rgb, 255:red, 0; green, 0; blue, 0 }  ][line width=0.75]      (0, 0) circle [x radius= 3.35, y radius= 3.35]   ;
\draw [shift={(78,96)}, rotate = 333.43] [color={rgb, 255:red, 0; green, 0; blue, 0 }  ][fill={rgb, 255:red, 0; green, 0; blue, 0 }  ][line width=0.75]      (0, 0) circle [x radius= 3.35, y radius= 3.35]   ;
\draw [line width=0.75]    (38,76) -- (118,116) ;
\draw [shift={(118,116)}, rotate = 26.57] [color={rgb, 255:red, 0; green, 0; blue, 0 }  ][fill={rgb, 255:red, 0; green, 0; blue, 0 }  ][line width=0.75]      (0, 0) circle [x radius= 3.35, y radius= 3.35]   ;
\draw [shift={(38,76)}, rotate = 26.57] [color={rgb, 255:red, 0; green, 0; blue, 0 }  ][fill={rgb, 255:red, 0; green, 0; blue, 0 }  ][line width=0.75]      (0, 0) circle [x radius= 3.35, y radius= 3.35]   ;
\draw [line width=0.75]  [dash pattern={on 3pt off 3pt}]  (118,36) -- (158,16) ;

\draw [line width=0.75]  [dash pattern={on 3pt off 3pt}]  (118,116) -- (158,136) ;

\draw [line width=0.75]  [dash pattern={on 3pt off 3pt}]  (118,76) -- (158,56) ;

\draw [line width=0.75]  [dash pattern={on 3pt off 3pt}]  (118,76) -- (158,96) ;

\draw [line width=0.75]  [dash pattern={on 3pt off 3pt}]  (118,116) -- (158,96) ;

\draw    (218,111) -- (218,23) (214,81) -- (222,81)(214,51) -- (222,51) ;
\draw [shift={(218,21)}, rotate = 450] [fill={rgb, 255:red, 0; green, 0; blue, 0 }  ][line width=0.75]  [draw opacity=0] (8.93,-4.29) -- (0,0) -- (8.93,4.29) -- cycle    ;

\draw    (218,111) -- (376,111) ;
\draw [shift={(378,111)}, rotate = 180] [fill={rgb, 255:red, 0; green, 0; blue, 0 }  ][line width=0.75]  [draw opacity=0] (8.93,-4.29) -- (0,0) -- (8.93,4.29) -- cycle    ;

\draw [line width=0.75]    (218,111) -- (278,51) ;
\draw [shift={(278,51)}, rotate = 315] [color={rgb, 255:red, 0; green, 0; blue, 0 }  ][fill={rgb, 255:red, 0; green, 0; blue, 0 }  ][line width=0.75]      (0, 0) circle [x radius= 3.35, y radius= 3.35]   ;
\draw [shift={(218,111)}, rotate = 315] [color={rgb, 255:red, 0; green, 0; blue, 0 }  ][fill={rgb, 255:red, 0; green, 0; blue, 0 }  ][line width=0.75]      (0, 0) circle [x radius= 3.35, y radius= 3.35]   ;
\draw [line width=0.75]    (278,81) -- (248,81) ;
\draw [shift={(248,81)}, rotate = 180] [color={rgb, 255:red, 0; green, 0; blue, 0 }  ][fill={rgb, 255:red, 0; green, 0; blue, 0 }  ][line width=0.75]      (0, 0) circle [x radius= 3.35, y radius= 3.35]   ;
\draw [shift={(278,81)}, rotate = 180] [color={rgb, 255:red, 0; green, 0; blue, 0 }  ][fill={rgb, 255:red, 0; green, 0; blue, 0 }  ][line width=0.75]      (0, 0) circle [x radius= 3.35, y radius= 3.35]   ;
\draw [line width=0.75]  [dash pattern={on 3pt off 3pt}]  (308,51) -- (278,51) ;

\draw [line width=0.75]    (248,111) -- (308,51) ;
\draw [shift={(308,51)}, rotate = 315] [color={rgb, 255:red, 0; green, 0; blue, 0 }  ][fill={rgb, 255:red, 0; green, 0; blue, 0 }  ][line width=0.75]      (0, 0) circle [x radius= 3.35, y radius= 3.35]   ;
\draw [shift={(248,111)}, rotate = 315] [color={rgb, 255:red, 0; green, 0; blue, 0 }  ][fill={rgb, 255:red, 0; green, 0; blue, 0 }  ][line width=0.75]      (0, 0) circle [x radius= 3.35, y radius= 3.35]   ;
\draw [line width=0.75]    (308,81) -- (278,81) ;
\draw [shift={(278,81)}, rotate = 180] [color={rgb, 255:red, 0; green, 0; blue, 0 }  ][fill={rgb, 255:red, 0; green, 0; blue, 0 }  ][line width=0.75]      (0, 0) circle [x radius= 3.35, y radius= 3.35]   ;
\draw [shift={(308,81)}, rotate = 180] [color={rgb, 255:red, 0; green, 0; blue, 0 }  ][fill={rgb, 255:red, 0; green, 0; blue, 0 }  ][line width=0.75]      (0, 0) circle [x radius= 3.35, y radius= 3.35]   ;
\draw [line width=0.75]    (278,111) -- (338,51) ;
\draw [shift={(338,51)}, rotate = 315] [color={rgb, 255:red, 0; green, 0; blue, 0 }  ][fill={rgb, 255:red, 0; green, 0; blue, 0 }  ][line width=0.75]      (0, 0) circle [x radius= 3.35, y radius= 3.35]   ;
\draw [shift={(278,111)}, rotate = 315] [color={rgb, 255:red, 0; green, 0; blue, 0 }  ][fill={rgb, 255:red, 0; green, 0; blue, 0 }  ][line width=0.75]      (0, 0) circle [x radius= 3.35, y radius= 3.35]   ;
\draw [line width=0.75]  [dash pattern={on 3pt off 3pt}]  (338,51) -- (308,51) ;

\draw [line width=0.75]  [dash pattern={on 3pt off 3pt}]  (308,21) -- (278,51) ;

\draw [line width=0.75]  [dash pattern={on 3pt off 3pt}]  (338,21) -- (308,51) ;

\draw [line width=0.75]  [dash pattern={on 3pt off 3pt}]  (368,21) -- (338,51) ;

\draw [line width=0.75]  [dash pattern={on 3pt off 3pt}]  (368,51) -- (338,51) ;

\draw [line width=0.75]    (308,111) -- (278,111) ;
\draw [shift={(278,111)}, rotate = 180] [color={rgb, 255:red, 0; green, 0; blue, 0 }  ][fill={rgb, 255:red, 0; green, 0; blue, 0 }  ][line width=0.75]      (0, 0) circle [x radius= 3.35, y radius= 3.35]   ;
\draw [shift={(308,111)}, rotate = 180] [color={rgb, 255:red, 0; green, 0; blue, 0 }  ][fill={rgb, 255:red, 0; green, 0; blue, 0 }  ][line width=0.75]      (0, 0) circle [x radius= 3.35, y radius= 3.35]   ;
\draw [line width=0.75]    (338,81) -- (308,81) ;
\draw [shift={(308,81)}, rotate = 180] [color={rgb, 255:red, 0; green, 0; blue, 0 }  ][fill={rgb, 255:red, 0; green, 0; blue, 0 }  ][line width=0.75]      (0, 0) circle [x radius= 3.35, y radius= 3.35]   ;
\draw [shift={(338,81)}, rotate = 180] [color={rgb, 255:red, 0; green, 0; blue, 0 }  ][fill={rgb, 255:red, 0; green, 0; blue, 0 }  ][line width=0.75]      (0, 0) circle [x radius= 3.35, y radius= 3.35]   ;
\draw [line width=0.75]  [dash pattern={on 3pt off 3pt}]  (368,51) -- (338,81) ;

\draw [line width=0.75]  [dash pattern={on 3pt off 3pt}]  (338,81) -- (308,111) ;

\draw (64.5,20) node   {$\phi ( t)$};
\draw (174,91) node   {$t$};
\draw (243.5,26) node   {$\psi ( t)$};
\draw (17.5,57) node   {$\frac{\pi }{2}$};
\draw (373,126) node   {$t$};
\draw (202.5,81) node   {$\pi $};
\draw (97.5,145) node  [align=left] {{\fontfamily{ptm}\selectfont {\large (a)}}};
\draw (287,145) node  [align=left] {{\fontfamily{ptm}\selectfont {\large (b)}}};

\end{tikzpicture}
\caption{CPM trellis (a) and its tilted version (b), $M_{\text{cpm}}=2$, $h=1/2$, $\phi_0=0$, $L_{\text{cpm}}=1$ and linear phase transition}
\label{fig:tilted_phase_msk}       
\end{center}
\vspace{-0.75em}
\end{figure}
In general, the corresponding phase trellis of \eqref{eq:cpm:phase_term} is time variant, which means that the possible phase states are time-dependent. In order to avoid the time-dependency, a time invariant trellis is constructed by tilting the phase according to the decomposition approach in \cite{Rimoldi_1988}. The tilt corresponds to a frequency offset applied to the CPM signal, i.e., the phase term becomes $\psi(t) = \phi(t)  + 2 \pi  \Delta f t $,
where $\Delta f = h (M_{\textrm{cpm}}   -1)/2  T_{\textrm{s}} $. In this context, Fig.~\ref{fig:tilted_phase_msk} shows the tilted version of a MSK signal. Taking into account the tilted trellis, a different symbol notation $x_k=  (\alpha_k  +  M_{\textrm{cpm}} - 1) / 2  $ can be considered, which then corresponds to the symbol alphabet $\mathcal{X} = \left\{  0, 1,\ldots  ,  M_{\textrm{cpm}}-1   \right\} $. The tilted CPM phase $\psi(t)$ within one symbol interval with duration $T_{\textrm{s}}$, letting $t=\tau+kT_{\textrm{s}}$, can be fully described by the state definition $\tilde{s}_{k} = \left[  \beta_{k-L_{\textrm{cpm}}} , x^{k}_{k-L_{\textrm{cpm}}+1}   \right]$ in terms of 
\begin{align}
\psi (\tau  +  k T_{\textrm{s}} ) =  &   \frac{2 \pi}{P_{\textrm{cpm}}} \beta_{k-L_{\textrm{cpm}}} \label{eq:cpm:Rimoldi}\\
& + 2 \pi h \sum_{l=0}^{L_{\textrm{cpm}}-1} \left( 2 x_{k-l} - M_{\textrm{cpm}} + 1\right) f(\tau+l T_{\textrm{s}}) \notag \\   
& + \pi h \left( M_{\textrm{cpm}} -1 \right) \left( \frac{\tau}{T_{\textrm{s}}} + L_{\textrm{cpm}} - 1 \right)  + \varphi_0 \text{,} \notag
\end{align}
where the absolute phase state
$\beta_{k-L_{\textrm{cpm}}}$ can be reduced to
\begin{align}
\beta_{k-L_{\textrm{cpm}}} = \left( K_{\textrm{cpm}}   \sum_{l=0}^{k-L_{\textrm{cpm}}}   x_l    \right) \bmod P_{\textrm{cpm}} \notag \text{,}
\end{align}
which is related to the $2\pi$-wrapped accumulated phase contributions of the input symbols that are prior to the CPM memory.

In the sequel a discrete time description is considered which implies that the CPM phase is represented in a vector notation. The corresponding tilted CPM phase $\psi (\tau  +  k T_{\textrm{s}} )$ for one symbol interval, i.e.,  $0 \leq \tau < T_{\textrm{s}}$, is then discretized into $MD$ samples, which composes the vector denoted by
$\boldsymbol{\psi}_{k}(\tilde{s}_{k})  =  [\psi(\frac{T_{\textrm{s}}}{MD}(kMD+1)),\psi(\frac{T_{\textrm{s}}}{MD}(kMD+2)),\dots,\psi(T_{\textrm{s}}(k+1))]^T$, where $M$ is the oversampling factor, and $D$ is a higher resolution multiplier. The tilt of the phase can be established in the actual communication system by receiving at an intermediate frequency (IF). With this, we can consider that different low-IF frequencies can be used, which motivates the definition of $\psi_{\textrm{IF}}(t) = \psi(t) + 2 \pi \frac{n_{\textrm{IF}}}{T_s} t $. Choosing $n_{\textrm{IF}} > 0$ is promising because the appearance of zero-crossings can be adjusted, as proposed in \cite{Bender_SPAWC2019}.
Nevertheless, in the considered examples
$n_{\textrm{IF}}=0$ is chosen for simplicity.

Regarding the discrete system model in Fig.~\ref{fig:system_model}, the CPM modulator takes the input sequence $x^n$ and generates the transmit signal $ \sqrt{ \frac{ E_s}{ T_{\textrm{s}} }}  e^{j \boldsymbol{\psi}_{k}(\tilde{s}_{k}) }$, where $E_s$ is the symbol energy, i.e., it already takes into account the frequency offset.

\subsubsection{Receive filter and 1-bit quantization}

The receive filter $g(t)$ has an impulse response of length $T_g$. In the discrete model for expressing a subsequence of $(\eta +1)$ oversampling output symbols it is represented in a matrix form with $\boldsymbol{G}$, as a $MD(\eta+1) \times MD(L_g+\eta+1)$ Toepliz matrix, as  described in equation (17) in \cite{Landau_CPM_2018}, whose first row is $[\boldsymbol{g}^T,\boldsymbol{0}_{MD(\eta+1)}^T]$, where $\boldsymbol{g}^T=[g( L_g T_{\textrm{s}} ), g( \frac{T_{\textrm{s}}}{MD} (L_gMD-1)),\dots,g(\frac{T_{\textrm{s}}}{MD})]$.
A higher sampling grid in the waveform signal, in the noise generation and in the filtering is adopted to adequately model the aliasing effect. This receive filtering yields an increase of memory in the system by $L_g$ symbols, where $(L_{g}-1) T_{\textrm{s}} < T_g \leq L_{g} T_{\textrm{s}}$. This motivates the definition of the overall memory in terms of $L=L_{\textrm{cpm}}+L_g$.

The filtered samples are decimated to the vector $\boldsymbol{z}^{k}_{k-\eta}$ according to the oversampling factor $M$, by multiplication with the $D$-fold decimation matrix $\boldsymbol{D}$, as described in equation (16) in \cite{Landau_CPM_2018}, with dimensions $M(\eta+1) \times MD(\eta+1)$. Then, the result $\boldsymbol{z}^{k}_{k-\eta}$ is 1-bit quantized to the vector $\boldsymbol{y}^{k}_{k-\eta}$. These operations can be represented by the following equations
\begin{align}
\boldsymbol{y}^{k}_{k-\eta} = Q \left( \boldsymbol{z}^{k}_{k-\eta}  \right) =  Q  \left(  \boldsymbol{D} \  \boldsymbol{G} \left[ \sqrt{  \frac{E_{\textrm{s}}}{T_{\textrm{s}}} }   e^{\boldsymbol{\psi}^{k}_{k-\eta-L_g}}    +  \boldsymbol{n}^{k}_{k-\eta-L_g }  \right] \right)   \text{,}
\label{eq:cpm_received_signal}
\end{align}
where the quantization operator $Q(\cdot)$ is applied element-wise. The quantization of $\boldsymbol{z}_k$ is described by ${y}_{k,m} =\mathrm{sgn}(\mathrm{Re}\left\{{z}_{k,m}\right\})+j \mathrm{sgn}(\mathrm{Im}\left\{{z}_{k,m}\right\})$, where $m$ denotes the oversampling index which runs from $1$ to $M$ and ${y}_{k,m} \in \left\{ 1+j,1-j,-1+j,-1-j \right\}$. The vector $\boldsymbol{n}^{k}_{k-\eta-L_g}$ contains complex zero-mean white Gaussian noise samples with variance $\sigma_n^2=N_0$.

\section{Proposed Faster-than-Nyquist CPM Waveform}
\label{sec:proposed_wf}

As known from linear modulation schemes, a faster-than-Nyquist signaling can yield a benefit for the design of zero-crossings \cite{Landau_ONEBIT_2018}, which is key for channels with 1-bit quantization at the receiver.
 In this section, a new subclass of CPM waveforms is introduced which provides relatively high, signaling rate and
 high spectral efficiency at the same time. Illustrated special configurations of the proposed waveform only require low-complexity at the transmitter and receiver.

In the sequel the proposed waveform is considered with the rectangular frequency pulse
with duration $T_{\textrm{cpm}}$, but the extension to frequency pulses like raised cosine and or Gaussian pulses would be straight forward.
\begin{figure}[t]
\begin{center}
\captionsetup{justification=centering,font=scriptsize}
\input{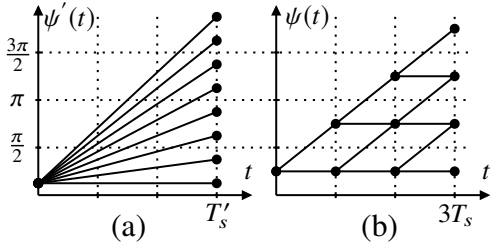}
\caption{8-symbol CPFSK (a) and three-symbol-period FTN-CPM (b) tilted trellises }
\label{fig:eight_symbol_wf}       
\end{center}
\vspace{-0.5em}
\end{figure}
\begin{figure}[t]
\begin{center}
\captionsetup{justification=centering,font=scriptsize}
 \includegraphics[scale=0.27]{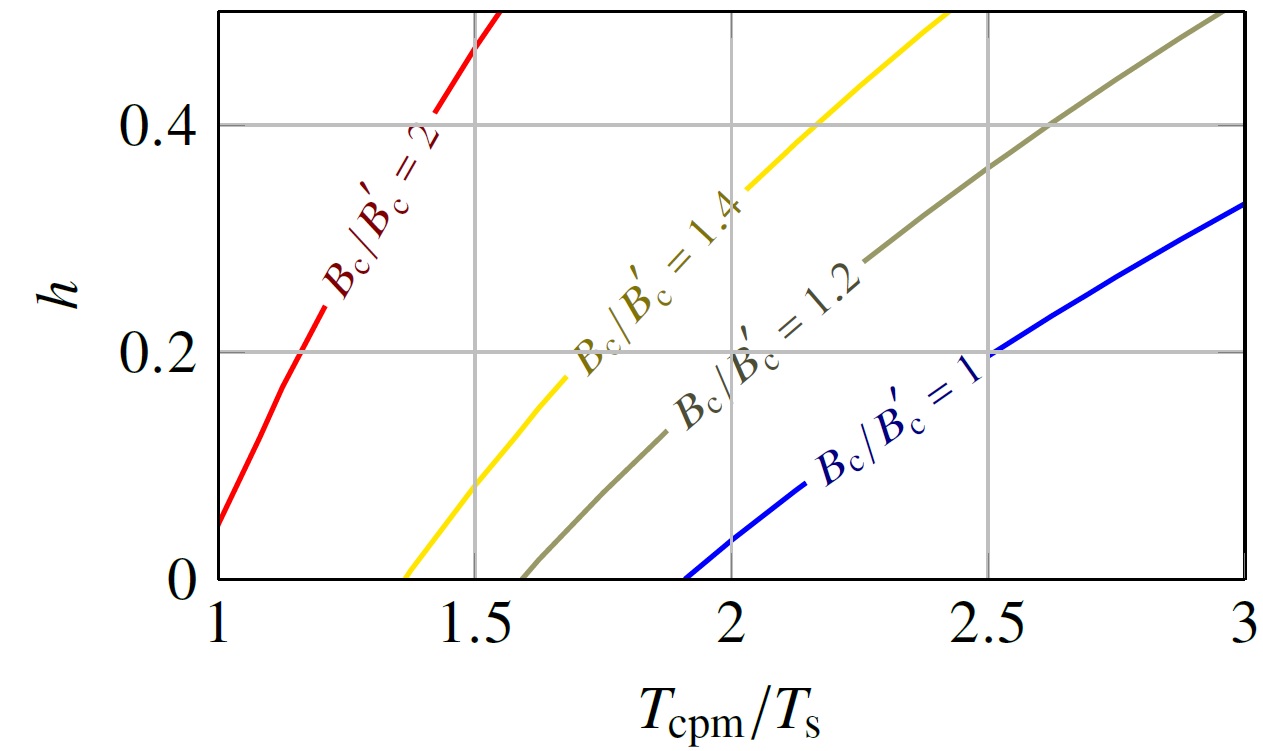}
\caption{ Equi-bandwidth ($B_{\text{c}} / B_{\text{c}}^{'} $) lines due to Carson's criterion,  $T_{\text{s}}^{'} / T_{\text{s}}=3$, $M_{\textrm{cpm}}=2$   }
\label{fig:carson_ratio}
\end{center}
\vspace{-0.5em}
\end{figure}

\subsection{Proof-of-concept of 
Faster-than-Nyquist CPM}

In this section, it is shown that CPM signals can be constructed with significantly reduced symbol duration as compared to standard CPFSK, which convey the same information per time interval and occupy the same bandwidth.
Because of its mathematical tractability, Carson's bandwidth criterion, as used in \cite{Kuo_2004,Barbieri_2007}, is considered in this section. 
Following the steps in \cite{Kuo_2004},
Carson's bandwidth  
for CPM signals with i.u.d.\ input and rectangular frequency pulse can be expressed as
\begin{align}
    B_{\textrm{c}} = h \sqrt{ (M_{\textrm{cpm}}^2-1) ( 3 T_{\textrm{s}} T_{\textrm{cpm}} )^{-1} } + T_{\textrm{cpm}}^{-1} \text{.}
\end{align}
As a reference, a standard CPFSK signal shall be considered, whose parameters are indicated with $'$. The reference CPFSK is fully described by $T_{\text{s}}^{'}$, $M_{\textrm{cpm}}^{'}$, $h^{'}=\frac{1}{M_{\textrm{cpm}}^{'}}$, $T_{\textrm{cpm}}^{'}=T_{\text{s}}^{'}$ and $B_{\textrm{c}}^{'}$. 
Now, it is aimed to construct a FTN CPM signal with a shorter symbol duration $T_{\text{s}}$, such that the ratio $T_{\text{s}}^{'} / T_{\text{s}}$ is an integer value, as shown in Fig.~\ref{fig:eight_symbol_wf}.
The conveyed information per time interval is equal for both signals by defining  $M_{\textrm{cpm}}^{T_{\text{s}}^{'} / T_{\text{s}}}=M_{\textrm{cpm}}^{'}$.
With this, the relation between the bandwidth of the FTN-CPM signal and the reference CPFSK signal can be expressed as
\begin{align}
\frac{B_{\textrm{c}}}{B_{\textrm{c}}^{'}}=  \frac{T_{\text{s}}^{'}}{T_{\text{s}}} \left( h \sqrt{ \frac{ M_{\textrm{cpm}}^2 -1 } {3 T_{\textrm{cpm}} / T_{\textrm{s}}}}  + \frac{T_{\textrm{s}}}{T_{\textrm{cpm}}} \right)
\left( 1+ \frac{1}{M_{\textrm{cpm}}^{T_{\text{s}}^{'} / T_{\textrm{s}}}} \sqrt{  \frac{M_{\textrm{cpm}}^{2 T_{\text{s}}^{'} / T_{\text{s}}}  -1}{3}} \right) ^{-1}
\text{.}
\label{eq:relative_Bc}
\end{align}
For the case of predefined design parameter
$\frac{T_{\text{s}}^{'}}{T_{\text{s}}}$ and $M_{\textrm{cpm}}$, the relation in \eqref{eq:relative_Bc} is a function of modulation index $h$ and relative frequency pulse length $T_{\textrm{cpm}}/ T_{\textrm{s}}$.
Aiming for a high spectral efficiency for the FTN-CPM signal a low relative bandwidth \eqref{eq:relative_Bc} is promising. An example case is illustrated for $T_{\text{s}}^{'} / T_{\text{s}}=3$ and $M_{\textrm{cpm}}=2$. As can be seen, 
the bandwidth increase brought by the higher signaling rate can be compensated by adjustment of the modulation index $h$ and the length of the frequency pulse $T_{\textrm{cpm}}$.
In the sequel the FTN-CPM waveform configurations are detailed which are promising in the presence of 1-bit quantization at the receiver.

\subsection{FTN-CPM for 1-bit quantization at the receiver}

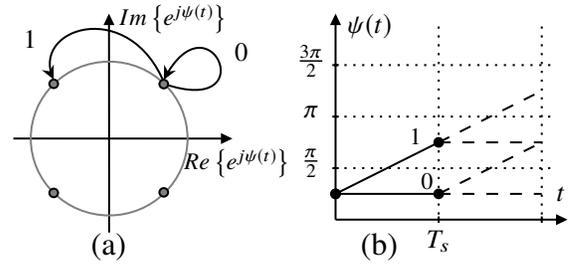
\begin{figure}[t]
\begin{center}
\captionsetup{justification=centering,font=scriptsize}
\tikzset{every picture/.style={line width=0.75pt}} 

\begin{tikzpicture}[x=0.75pt,y=0.75pt,yscale=-0.65,xscale=0.65]

\draw    (124.34,190) -- (124.19,22) ;
\draw [shift={(124.19,20)}, rotate = 449.95] [fill={rgb, 255:red, 0; green, 0; blue, 0 }  ][line width=0.75]  [draw opacity=0] (8.93,-4.29) -- (0,0) -- (8.93,4.29) -- cycle    ;

\draw    (50,117.02) -- (218,117.06) ;
\draw [shift={(220,117.06)}, rotate = 180.01] [fill={rgb, 255:red, 0; green, 0; blue, 0 }  ][line width=0.75]  [draw opacity=0] (8.93,-4.29) -- (0,0) -- (8.93,4.29) -- cycle    ;

\draw  [color={rgb, 255:red, 128; green, 128; blue, 128 }  ,draw opacity=1 ] (63.52,117.02) .. controls (63.52,84.04) and (90.75,57.31) .. (124.34,57.31) .. controls (157.93,57.31) and (185.17,84.04) .. (185.17,117.02) .. controls (185.17,150) and (157.93,176.73) .. (124.34,176.73) .. controls (90.75,176.73) and (63.52,150) .. (63.52,117.02) -- cycle ;
\draw    (167.38,74.75) .. controls (140.17,7.12) and (64.35,30.56) .. (80.51,69.45) ;
\draw [shift={(81.31,71.24)}, rotate = 244.23000000000002] [fill={rgb, 255:red, 0; green, 0; blue, 0 }  ][line width=0.75]  [draw opacity=0] (10.72,-5.15) -- (0,0) -- (10.72,5.15) -- (7.12,0) -- cycle    ;

\draw    (167.38,74.75) .. controls (244.79,100.83) and (202.36,3.75) .. (167.23,70.42) ;
\draw [shift={(166.7,71.44)}, rotate = 297.15999999999997] [fill={rgb, 255:red, 0; green, 0; blue, 0 }  ][line width=0.75]  [draw opacity=0] (10.72,-5.15) -- (0,0) -- (10.72,5.15) -- (7.12,0) -- cycle    ;

\draw  [fill={rgb, 255:red, 128; green, 128; blue, 128 }  ,fill opacity=1 ] (77.93,74.56) .. controls (77.93,72.72) and (79.45,71.24) .. (81.31,71.24) .. controls (83.18,71.24) and (84.69,72.72) .. (84.69,74.56) .. controls (84.69,76.39) and (83.18,77.87) .. (81.31,77.87) .. controls (79.45,77.87) and (77.93,76.39) .. (77.93,74.56) -- cycle ;
\draw  [fill={rgb, 255:red, 128; green, 128; blue, 128 }  ,fill opacity=1 ] (163.32,74.75) .. controls (163.32,72.92) and (164.84,71.44) .. (166.7,71.44) .. controls (168.57,71.44) and (170.08,72.92) .. (170.08,74.75) .. controls (170.08,76.59) and (168.57,78.07) .. (166.7,78.07) .. controls (164.84,78.07) and (163.32,76.59) .. (163.32,74.75) -- cycle ;
\draw  [fill={rgb, 255:red, 128; green, 128; blue, 128 }  ,fill opacity=1 ] (77.93,159.26) .. controls (77.93,157.43) and (79.45,155.94) .. (81.31,155.94) .. controls (83.18,155.94) and (84.69,157.43) .. (84.69,159.26) .. controls (84.69,161.09) and (83.18,162.58) .. (81.31,162.58) .. controls (79.45,162.58) and (77.93,161.09) .. (77.93,159.26) -- cycle ;
\draw  [fill={rgb, 255:red, 128; green, 128; blue, 128 }  ,fill opacity=1 ] (163.31,158.82) .. controls (163.31,156.98) and (164.83,155.5) .. (166.69,155.5) .. controls (168.56,155.5) and (170.07,156.98) .. (170.07,158.82) .. controls (170.07,160.65) and (168.56,162.13) .. (166.69,162.13) .. controls (164.83,162.13) and (163.31,160.65) .. (163.31,158.82) -- cycle ;
\draw [color={rgb, 255:red, 0; green, 0; blue, 0 }  ,draw opacity=1 ] [dash pattern={on 0.84pt off 2.51pt}]  (460,30) -- (460,180) ;

\draw [color={rgb, 255:red, 0; green, 0; blue, 0 }  ,draw opacity=1 ] [dash pattern={on 0.84pt off 2.51pt}]  (380,30) -- (380,180) ;

\draw    (300,180) -- (300,22) (297.5,140) -- (302.5,140)(297.5,100) -- (302.5,100)(297.5,60) -- (302.5,60) ;
\draw [shift={(300,20)}, rotate = 450] [fill={rgb, 255:red, 0; green, 0; blue, 0 }  ][line width=0.75]  [draw opacity=0] (8.93,-4.29) -- (0,0) -- (8.93,4.29) -- cycle    ;

\draw    (300,180) -- (478,180) (380,177.5) -- (380,182.5)(460,177.5) -- (460,182.5) ;
\draw [shift={(480,180)}, rotate = 180] [fill={rgb, 255:red, 0; green, 0; blue, 0 }  ][line width=0.75]  [draw opacity=0] (8.93,-4.29) -- (0,0) -- (8.93,4.29) -- cycle    ;

\draw [line width=0.75]    (300,160) -- (380,120) ;
\draw [shift={(380,120)}, rotate = 333.43] [color={rgb, 255:red, 0; green, 0; blue, 0 }  ][fill={rgb, 255:red, 0; green, 0; blue, 0 }  ][line width=0.75]      (0, 0) circle [x radius= 3.35, y radius= 3.35]   ;
\draw [shift={(300,160)}, rotate = 333.43] [color={rgb, 255:red, 0; green, 0; blue, 0 }  ][fill={rgb, 255:red, 0; green, 0; blue, 0 }  ][line width=0.75]      (0, 0) circle [x radius= 3.35, y radius= 3.35]   ;
\draw [color={rgb, 255:red, 0; green, 0; blue, 0 }  ,draw opacity=1 ] [dash pattern={on 0.84pt off 2.51pt}]  (300,140) -- (470,140) ;

\draw [color={rgb, 255:red, 0; green, 0; blue, 0 }  ,draw opacity=1 ] [dash pattern={on 0.84pt off 2.51pt}]  (300,100) -- (470,100) ;

\draw [color={rgb, 255:red, 0; green, 0; blue, 0 }  ,draw opacity=1 ] [dash pattern={on 0.84pt off 2.51pt}]  (300,60) -- (470,60) ;

\draw [line width=0.75]    (300,160) -- (380,160) ;
\draw [shift={(380,160)}, rotate = 0] [color={rgb, 255:red, 0; green, 0; blue, 0 }  ][fill={rgb, 255:red, 0; green, 0; blue, 0 }  ][line width=0.75]      (0, 0) circle [x radius= 3.35, y radius= 3.35]   ;
\draw [shift={(300,160)}, rotate = 0] [color={rgb, 255:red, 0; green, 0; blue, 0 }  ][fill={rgb, 255:red, 0; green, 0; blue, 0 }  ][line width=0.75]      (0, 0) circle [x radius= 3.35, y radius= 3.35]   ;
\draw [line width=0.75]  [dash pattern={on 4.5pt off 4.5pt}]  (380,120) -- (460,80) ;

\draw [line width=0.75]  [dash pattern={on 4.5pt off 4.5pt}]  (380,120) -- (460,120) ;

\draw [line width=0.75]  [dash pattern={on 4.5pt off 4.5pt}]  (380,160) -- (460,160) ;

\draw [line width=0.75]  [dash pattern={on 4.5pt off 4.5pt}]  (380,160) -- (460,120) ;

\draw (223,136) node [scale=0.9]  {$Re\left\{e^{j\psi ( t)}\right\}$};
\draw (173.5,24) node [scale=0.9]  {$Im\left\{e^{j\psi ( t)}\right\}$};
\draw (227,50) node   {$0$};
\draw (63,40) node   {$1$};
\draw (281,136) node   {$\frac{\pi }{2}$};
\draw (326.5,30) node   {$\psi ( t)$};
\draw (475,160) node   {$t$};
\draw (124,201) node  [align=left] {{\fontfamily{ptm}\selectfont {\large (a)}}};
\draw (380.5,197) node   {$T_{s}$};
\draw (281,56.5) node   {$\frac{3\pi }{2}$};
\draw (281,96) node   {$\pi $};
\draw (334,201) node  [align=left] {{\fontfamily{ptm}\selectfont {\large (b)}}};
\draw (371,150) node [scale=0.9]  {$0$};
\draw (362,115) node [scale=0.9]  {$1$};

\end{tikzpicture}
\caption{Tilted CPM constellation diagram (a) and trellis (b) of the proposed FTN-CPM with $T_{\text{cpm}}=T_{\text{s}}$, $h=1/4$ and $\phi_0=\pi/4$}
\label{fig:tilted_phase}       
\end{center}
\vspace{-0.5em}
\end{figure}

A widely used waveform design criterion for channels with 1-bit quantization at the receiver is given by the maximization of distance to the decision threshold \cite{Landau_SCC2013}.
By assuming that the receive filter $g(t)$ only marginally changes the signal phase $\psi(t)$ at the receiver, zero-crossings appear whenever the phase crosses integer multiples of $\frac{\pi}{2}$.
Considering that sampling rate is equal to the FTN signaling rate, the illustrated FTN-CPM phase tree on the RHS of Fig.~\ref{fig:eight_symbol_wf} is optimal in terms of distance to decision threshold. 
The corresponding binary FTN-CPM constellation diagram is shown in Fig.~\ref{fig:tilted_phase}, which implies that a zero-crossing conveys the transmit symbol 1 and 0 else.  
In order to achieve a spectral efficiency similar to the corresponding conventional CPFSK waveform, the length of the frequency pulse $T_{\textrm{cpm}}$ can be increased, cf.\ Fig.~\ref{fig:carson_ratio}, where different cases are examined in the sequel. 

\section{CPM Demodulation}
\label{sec:demodulation}
This section describes two demodulation methods.

\subsection{MAP Detection}
The MAP decision metric for each symbol $x_k$ is given by the A Posteriori Probability (APP) $P \left( x_{k} \vert \boldsymbol{y}^n \right) $.
For the considered system,  approximated APPs $P_{\textrm{aux}} \left( x_{k} \vert \boldsymbol{y}^n \right) $ can be computed via a BCJR algorithm \cite{BCJR_1974} that is based on an auxiliary channel law.
Depending on the receive filter, noise samples are correlated which then implies dependency on previous channel outputs, such that the channel law has the form $P(\boldsymbol{y}_k \vert \boldsymbol{y}^{k-1}, x^n )$.
In this case the consideration of an auxiliary channel law $W(\cdot)$ is required \cite{Landau_CPM_2018}, which can be described by
\begin{align}
W(\boldsymbol{y}_k \vert \boldsymbol{y}^{k-1}, x^n ) & =
P(\boldsymbol{y}_k \vert \boldsymbol{y}^{k-1}_{k-N}, \beta_{k-L-N} , x^{k}_{k-L-N+1}) \notag \\
& =\frac{
P( \boldsymbol{y}^k_{k-N}   \vert \beta_{k-L-N}, x^{k}_{k-L-N+1})}{P(  \boldsymbol{y}^{k-1}_{k-N} \vert \beta_{k-L-N}, x^{k-1}_{k-L-N+1})}
\text{,}
\label{eq:aux_channel}
\end{align}
where the dependency on $N$ previous channel outputs is taken into account.
With this, an extended state representation is required which is denoted by
\begin{align}
s_k=\begin{cases}
  [ \beta_{k-L-N+1} { ,} x^{k}_{k-L-N+2}  ] { ,}  & \text{ if } L+N> 1{ ,}   \\  
  [ \beta_{k} ] , & \text{ if }   L+N=1  \text{,}
\end{cases}
\label{eq:states_CPM}
\end{align}
where $L+N$ is the total memory of the system.
Based on the state notation
in \eqref{eq:states_CPM} the probabilities for the channel law can be cast as
$P\left(   \boldsymbol{y}^{k}_{k-N} \vert s_k, s_{k-1} \right)$ and
$P\left(   \boldsymbol{y}^{k-1}_{k-N} \vert s_{k-1} \right)$.
The auxiliary channel law probabilities \eqref{eq:aux_channel} 
involve a multivariate Gaussian integration in terms of
\begin{align}
 P\left(   \boldsymbol{y}^{k}_{k-N} \vert s_k, s_{k-1} \right) = 
 \int\limits_{ \boldsymbol{z}_{k-N}^{k} \in  \mathbb{Y}_{k-N}^{k}}  p( \boldsymbol{z}_{k-N}^{k}  \vert s_k, s_{k-1}    )  d \boldsymbol{z}_{k-N}^{k}  \text{,}
\label{eq:multivariate_integration}
\end{align}
where $\boldsymbol{z}_{k-N}^{k}$ is a complex Gaussian random vector that describes the input of the ADC, with a mean vector defined by $\boldsymbol{\mu}_{x} = \boldsymbol{D}\  \boldsymbol{G} \left[ \sqrt{  \frac{E_{\textrm{s}}}{T_{\textrm{s}}} }   e^{\boldsymbol{\psi}^{k}_{k-N-L_g}} \right]$, and covariance matrix $\boldsymbol{R}=\sigma_n^2 \boldsymbol{D}\boldsymbol{G}\boldsymbol{G}^H\boldsymbol{D}^T$, with $\boldsymbol{D}$ and $\boldsymbol{G}$ as introduced before with $\eta=N$.
 The integration interval is expressed in terms of the quantization region $ \mathbb{Y}_{k-N}^{k}$
that belongs to the channel output symbol $\boldsymbol{y}^{k}_{k-N}$.
After rewriting \eqref{eq:multivariate_integration} as a real valued multivariate Gaussian integration, how it is done in equation (21) in \cite{Landau_CPM_2018}, the algorithm in \cite{Genz_1992} can be applied.
Finally the BCJR algorithm provides the
probabilities $P_{\textrm{aux}} \left( s_{k-1},s_{k} \vert \boldsymbol{y}^n \right)$
which are subsequently used for computing the symbol APPs via
$P_{\textrm{aux}} \left( x_{k} \vert \boldsymbol{y}^n \right) = \sum_{\forall s_{k-1},s_{k} \supseteq x_{k}} P_{\textrm{aux}} \left( s_{k-1},s_{k} \vert \boldsymbol{y}^n \right)$.
The multivariate Gaussian integration \eqref{eq:multivariate_integration} becomes computationally expensive when $M$, $M_{\textrm{cpm}}$ and the channel memory are high, as detailed in \cite{Bender_SPAWC2019}. 
Note that for uncorrelated noise samples, subsequent channel outputs are independent, such that $P(\boldsymbol{y}_k \vert \boldsymbol{y}^{k-1}, x^n )=
P(\boldsymbol{y}_k \vert  x^n )
$, obviating the need for an auxiliary channel law.

\subsection{The Proposed Simple Demodulator}

\begin{figure}[t]
\begin{center}
\captionsetup{justification=centering,font=scriptsize}
\input{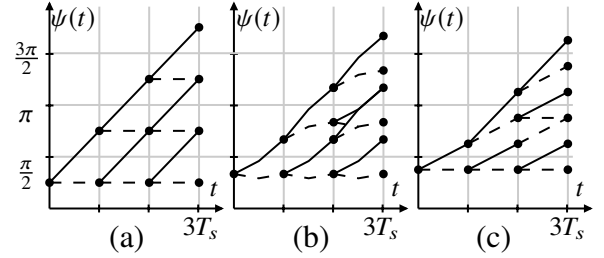}
\caption{Simple receive strategy: decide for $0$ (dashed line) and 1 (solid line); Different FTN-CPM configurations are shown: (a)  $T_{\text{cpm}}=T_{\text{s}}$, (b) $T_{\text{cpm}}=1.5 T_{\text{s}}$ and (c) $T_{\text{cpm}}=2 T_{\text{s}}$}
\label{fig:simple_rx}       
\end{center}
\vspace{-1em}
\end{figure}

In order to relieve the computational load at the receiver, 
some versions of the
proposed FTN-CPM scheme can be demodulated with an alternative simple strategy.
 The receive strategy for the binary FTN-CPM case with $h=\frac{1}{4}$, and sufficiently small $T_{\textrm{cpm}}$, like $T_{\textrm{cpm}}=2T_{\textrm{s}}$, only involves the evaluation of a change in real or imaginary part, depending on the previous sample $y_{k-1}$, which can be cast as
 \begin{align}
 \hat{x}_k=\begin{cases}
  \frac{1}{2} \vert  \{    \mathrm{Re}\{ y_k  \} - \mathrm{Re}\{ y_{k-1}  \}     \vert  \text{,}
  & \text{ if } y_{k-1} \in \{  1+j, -1-j   \} \text{,}    \\  
   \frac{1}{2} \vert  \{    \mathrm{Im}\{ y_k  \} - \mathrm{Im}\{ y_{k-1}  \}    \vert  { ,}
  & \text{ if } y_{k-1} \in \{  1-j, -1+j   \} \text{.} 
\end{cases}\notag   
 \end{align}
Fig.~\ref{fig:simple_rx} illustrates the receiver decisions in a noise-free scenario 
for $T_{\text{cpm}}=T_{\textrm{s}}$ in Fig.~\ref{fig:simple_rx}(a),
$T_{\text{cpm}}=1.5T_{\textrm{s}}$ in Fig.~\ref{fig:simple_rx}(b)
and $T_{\text{cpm}}=2T_{\textrm{s}}$ in Fig.~\ref{fig:simple_rx}(c), where the phase distortion brought by the receive filter is neglected for illustration purpose.
Note that for larger values for $T_{\textrm{cpm}}$ the noise sensitivity increases.
A special case is given by $T_{\text{cpm}}=2 T_{\textrm{s}}$ which results in the same number of equidistant constellation points as the corresponding 8-symbol CPFSK. 

\section{Numerical Results}
\label{sec:numerical_results}

In order to preserve the transmit waveform and its zero-crossings, 
a suboptimal short bandpass receive filter is considered as follows
\begin{align}
g(t)  =    \sqrt{ \frac{1}{T_g}}   \mathrm{rect}\left(\frac{t- T_{g} / 2  }{T_g}\right) \cdot e^{j 2 \pi \Delta f (t-T_g / 2) } \textrm{,}
\end{align}
 which is similar to the integrate and dump receiver considered in \cite{Costa_1994}. Note that the common receiver based on a matched filter bank is hardware demanding and not compatible with the considered 1-bit approach.
 Table \ref{table:waveform_params} gathers the simulation parameters for all considered CPM waveforms. 4-CPFSK \cite{Landau_CPM_2018} serves a standard reference waveform, which provides reliable communication without additional coding when considering 1-bit quantization. 
 The same holds for
 8-CPFSK \cite{Bender_SPAWC2019} which serves as reference waveform that does not require additional coding for $M=5$ and optimized low-IF with $n_{\textrm{IF}}=0.25$.
The proposed FTN-CPM is represented by the running example from Section~\ref{sec:proposed_wf} and \ref{sec:demodulation}
 specified by $M_{\textrm{cpm}}=2$, $h=\frac{1}{4}$ and rectangular frequency pulse with different durations $T_{\textrm{cpm}}$. 
 Note that for the considered FTN-CPM schemes the receive filter is such that noise samples are uncorrelated and an auxiliary channel law (specified by $N$) is not required.
{\small
\begin{table}[h]
\centering
\captionsetup{font=scriptsize}
	\begin{tabular}{|c|c|}
	\hline
	Waveform & Simulation Parameters \\
	\hline
	4-CPFSK \cite{Landau_CPM_2018} & \begin{tabular}{c} 
$M_{\textrm{cpm}}=4$, $L_{\textrm{cpm}}=1$, $T_g=0.5T_{\textrm{s}}$,\\ 
$h=1/4$, $n_{\textrm{IF}}=0$, $\phi_0=\pi/4$, $N=0$\end{tabular} \\
	\hline
	8-CPFSK \cite{Bender_SPAWC2019} & \begin{tabular}{c} 
$M_{\textrm{cpm}}=8$, $L_{\textrm{cpm}}=1$, $M=5$, $T_g=0.5T_{\textrm{s}}$,\\ 
$h=1/8$, $n_{\textrm{IF}}=0.25$, $\phi_0=\pi/8$, $N=0$\end{tabular} \\
	\hline
	Proposed FTN-CPM & \begin{tabular}{c} 
$M_{\textrm{cpm}}=2$, $M=1$, $T_g=T_{\textrm{s}}$,\\ 
$h=1/4$, $n_{\textrm{IF}}=0$, $\phi_0=\pi/4$\end{tabular} \\
	\hline
	\end{tabular}
	\caption{Considered waveforms}
	\label{table:waveform_params}
	\vspace{-0.5em}
\end{table}
}

In the sequel the adjustable power containment bandwidth 
$B_{90\%}$ is considered, where we refer to $90\%$ power containment as default and use $95\%$ as an alternative for some cases.

The considered $\mathrm{SNR}$ is defined by
\begin{align}
\mathrm{SNR} =     \frac{  \lim_{T \to \infty} \frac{1}{T}   \int_{T}   \left| x\left(t\right) \right|^2   dt    }
{N_0  \ B_{90\%} } =
\frac{E_\textrm{s}}{N_0}  (T_{\textrm{s}}
B_{90\%})^{-1}
\textrm{,}
\end{align}
where $x(t)= \sqrt{ \frac{ E_\textrm{s}}{ T_\textrm{s} }}  e^{j \psi(t)}$ is the complex low-IF representation of the signal and the noise power density $N_0$ corresponds to the variance of the noise samples in 
\eqref{eq:cpm_received_signal}.

\subsection{Spectral Efficiency and Effective Oversampling Ratio}
\begin{figure}[t]
\begin{center}
\captionsetup{justification=centering,font=scriptsize}
\definecolor{mycolor1}{rgb}{0.85000,0.32500,0.09800}%
\definecolor{mycolor2}{rgb}{0.46600,0.67400,0.18800}%
\definecolor{mycolor3}{rgb}{0.00000,0.44700,0.74100}%

\begin{tikzpicture}

\begin{axis}[%
width=0.85\columnwidth,
height=0.5\columnwidth,
scale only axis,
xmin=-5,
xmax=30,
xlabel={SNR [dB]},
ymin=0,
ymax=4,
ylabel={Spectral Eff. [bit/s/Hz]},
xmajorgrids,
ymajorgrids,
yminorgrids,
legend style={at={(1,0)},anchor=south east,draw=black,fill=white,legend cell align=left,font=\tiny}
]

\addplot [smooth,color=blue,solid, every mark/.append style={solid,scale=1.6}, mark=triangle]
  table[row sep=crcr]{%
0.902309461455717	0.799014968560162\\
5.90230946145572	1.62235194184515\\
10.9023094614557	2.45975401880306\\
15.9023094614557	3.1966925981982\\
20.9023094614557	3.79193594617084\\
25.9023094614557	3.89253506769337\\
30.9023094614557	3.89255885138816\\
};
\addlegendentry{Proposed FTN-CPM, $M=1$, $T_{\textrm{cpm}}=2T_{\textrm{s}}$}

\addplot [smooth,color=blue,solid, every mark/.append style={solid,scale=1.6}, mark=diamond]
  table[row sep=crcr]{%
0.460313627111645	0.740958018877109\\
5.46031362711165	1.52561811368599\\
10.4603136271116	2.42265804928183\\
15.4603136271116	3.1792999883886\\
20.4603136271116	3.50789141335583\\
25.4603136271116	3.51589262580115\\
30.4603136271116	3.5158926301407\\
};
\addlegendentry{Proposed FTN-CPM, $M=1$, $T_{\textrm{cpm}}=1.6T_{\textrm{s}}$}

\addplot [smooth,color=black,dashed, every mark/.append style={solid, fill=gray,draw=black}, mark=*]
  table[row sep=crcr]{%
-4.37121725598317	0.341802159794814\\
0.628782744016834	0.892695127058381\\
5.62878274401683	1.78999986161703\\
10.6287827440168	2.58854992606978\\
15.6287827440168	3.11919086060243\\
20.6287827440168	3.38005366018774\\
25.6287827440168	3.45188922765936\\
30.6287827440168	3.46736474507407\\
};
\addlegendentry{8-CPFSK, $M=5$, \cite{Bender_SPAWC2019}}

\addplot [smooth,color=blue,solid, every mark/.append style={solid}, mark=square]
  table[row sep=crcr]{%
-0.432317157767807	0.61769644194971\\
4.56768284223219	1.34033456819911\\
9.56768284223219	2.21013770848043\\
14.5676828422322	2.76773886151881\\
19.5676828422322	2.86266641628916\\
24.5676828422322	2.86267816494677\\
29.5676828422322	2.86267816494677\\
};
\addlegendentry{Proposed FTN-CPM, $M=1$, $T_{\textrm{cpm}}=T_{\textrm{s}}$}

\addplot [smooth,color=black,dashed, every mark/.append style={solid}, mark=star]
  table[row sep=crcr]{%
-4.26586046360554	0.348863676823218\\
0.734139536394462	0.899635888134777\\
5.73413953639446	1.74088633915594\\
10.7341395363945	2.26854479351828\\
15.7341395363945	2.36594452674367\\
20.7341395363945	2.36836257729089\\
25.7341395363945	2.36836257938708\\
30.7341395363945	2.36836257938711\\
};
\addlegendentry{4-CPFSK, $M=4$
\cite{Landau_CPM_2018}}

\addplot [smooth,color=black,dashed, every mark/.append style={solid, fill=white}, mark=o]
  table[row sep=crcr]{%
-4.25037009794924	0.294519636162432\\
0.74962990205076	0.777761806656639\\
5.74962990205076	1.58139878900233\\
10.7496299020508	2.2142459914508\\
15.7496299020508	2.36916538627223\\
20.7496299020508	2.37681857309993\\
25.7496299020508	2.37681910876028\\
30.7496299020508	2.3768197318506\\
};
\addlegendentry{4-CPFSK, $M=2$ \cite{Landau_CPM_2018}}

\end{axis}
\end{tikzpicture}%
\caption{Spectral Efficiency with respect to the $90\%$ power containment bandwidth}
\label{fig:spec_eff_cpm}
\end{center}
\vspace{-1em}
\end{figure}
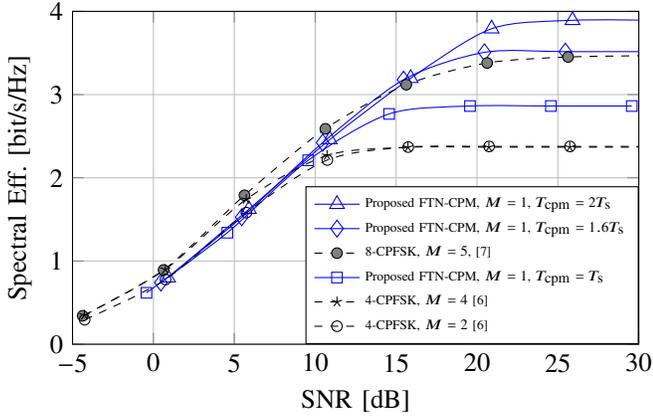

The spectral efficiency w.r.t.\ $B_{90\%}$ reads  
\begin{align}
\mathrm{spectral \ eff.} =  \frac{ I_{\textrm{bpcu}}  }  { B_{90\%} T_{\textrm{s}} }    \textrm{ [bit/s/Hz]}      \textrm{,}
\end{align}
where $I_{\textrm{bpcu}}$ is the achievable rate w.r.t.\ one symbol duration $T_{\textrm{s}}$, which is computed by applying the methods developed in \cite{Landau_CPM_2018}. The effective oversampling ratio w.r.t.\ $B_{90\%}$ is given by
\begin{align}
\mathrm{OSR}^{\prime}=    M   \left( B_{90\%} T_{\textrm{s}} \right)^{-1} \textrm{.}
\end{align}
Table \ref{table:bandtime_prod} displays computed values for effective oversampling factor and maximum spectral efficiency for the waveforms considered in Table \ref{table:waveform_params}. Moreover, Fig.~\ref{fig:spec_eff_cpm} illustrates the spectrum efficiency w.r.t.\ $B_{90\%}$ versus $\mathrm{SNR}$. 
For this bandwidth criterion choosing 
$T_{\textrm{cpm}}\geq1.6T_{\textrm{s}}$ can yield a higher spectral efficiency as the corresponding CPFSK waveform for medium and high SNR. When referring to the $B_{95\%}$ it is required to choose $T_{\textrm{cpm}} \geq 2T_{\textrm{s}}$ for approaching the spectral efficiency of the corresponding CPFSK, cf.\ Table~\ref{table:bandtime_prod}. 
{\small
\begin{table}[h]
\centering
\captionsetup{font=scriptsize}
	\begin{tabular}{|c|c|c|c|c|c|}
	\hline
	Waveform & $\frac{T_{\textrm{cpm}}}{T_{\textrm{s}}}$ & $M$ & $\frac{\log_2 M_{\textrm{cpm}}}{B_{90\%}T_{\textrm{s}}}$ & $\frac{\log_2 M_{\textrm{cpm}}}{B_{95\%}T_{\textrm{s}}}$ & $OSR^{\prime}$ \\
	\hline
	8-CPFSK \cite{Bender_SPAWC2019} & 1 & 5 & \bf{3.467} & \bf{2.873} & 5.778 \\
	4-CPFSK \cite{Landau_CPM_2018} & 1 & 4 & 2.372 & 1.976 & 4.744 \\
	4-CPFSK \cite{Landau_CPM_2018} & 1 & 2 & 2.372 & 1.976 & 2.372 \\
	 Proposed FTN-CPM & 1 & 1 & 2.853 & 1.983 & 2.853 \\
	 Proposed FTN-CPM & 1.2 & 1 & 3.079 & 2.176 & 3.079 \\
	 Proposed FTN-CPM & 1.4 & 1 & 3.297 & 2.359 & 3.297 \\
	 Proposed FTN-CPM & 1.6 & 1 & \bf{3.507} & 2.544 & 3.507 \\
	Proposed FTN-CPM & 1.8 & 1 & 3.691 & 2.720 & 3.691 \\
	 proposed FTN-CPM & 2 & 1 & 3.891 & \bf{2.881} & 3.891 \\
	\hline
	\end{tabular}
	\caption{Computed power containment bandwidths and effective oversampling factor}
	\label{table:bandtime_prod}
	\vspace{-1em}
\end{table}
}

\subsection{Bit Error Rate}
\begin{figure}[t]
\begin{center}
\captionsetup{justification=centering,font=scriptsize}
\begin{tikzpicture}

\begin{axis}[%
width=0.85\columnwidth,
height=0.5\columnwidth,
scale only axis,
grid style={color=gray!20},
xmin=0,
xmax=30,
xlabel={SNR [dB]},
ymode=log,
ymin=1e-06,
ymax=1,
yminorticks=true,
ylabel={BER},
xmajorgrids,
ymajorgrids,
yminorgrids,
legend entries={Proposed Simple Rx,BCJR},
legend style={at={(1,1)}, anchor=north east, legend cell align=left, align=left,font=\tiny}
]

\addlegendimage{solid,green, no marks}
\addlegendimage{dashed,black, no marks}

\addplot [smooth, solid, color=green, mark=square, mark options={solid, green}]
  table[row sep=crcr]{%
-0.455116757221515	0.4312\\
2.04488324277848	0.3835\\
4.54488324277848	0.310844827586207\\
7.04488324277848	0.219338028169014\\
9.54488324277849	0.117657971014493\\
12.0448832427785	0.0416236559139785\\
14.5448832427785	0.00794384236453202\\
17.0448832427785	0.000523867560430632\\
19.5448832427785	5.19393482449527e-06\\
20.7948832427785	4.14622348144565e-07\\
22.0448832427785	0\\
23.2948832427785	0\\
24.5448832427785	0\\
27.0448832427785	0\\
};

\addplot [smooth,dashed, color=black, mark=square, mark options={solid, black}]
  table[row sep=crcr]{%
-0.455116757221515	0.393\\
2.04488324277848	0.338083333333333\\
4.54488324277848	0.262879310344828\\
7.04488324277848	0.178154929577465\\
9.54488324277849	0.0969130434782609\\
12.0448832427785	0.0359784946236559\\
14.5448832427785	0.00737192118226601\\
17.0448832427785	0.00051249238269348\\
19.5448832427785	5.02638853983413e-06\\
20.7948832427785	4.14622348144565e-07\\
22.0448832427785	0\\
23.2948832427785	0\\
24.5448832427785	0\\
27.0448832427785	0\\
};

\addplot [smooth, solid, color=green, mark=diamond, mark options={solid, green,scale=1.6}]
  table[row sep=crcr]{%
0.439439088933854	0.4518\\
2.93943908893385	0.409769230769231\\
5.43943908893385	0.354720588235294\\
7.93943908893385	0.281231638418079\\
10.4394390889339	0.186081896551724\\
12.9394390889339	0.09866308835673\\
15.4394390889339	0.035421568627451\\
17.9394390889339	0.00673879331233341\\
20.4394390889339	0.000452933531377646\\
22.9394390889339	4.87249706583206e-06\\
24.1894390889339	2.18014838634955e-07\\
25.4394390889339	0\\
27.9394390889339	0\\
};

\addplot [smooth,dashed ,color=black, mark=diamond, mark options={solid, black,scale=1.6}]
  table[row sep=crcr]{%
0.439439088933854	0.4068\\
2.93943908893385	0.345846153846154\\
5.43943908893385	0.285632352941176\\
7.93943908893385	0.218067796610169\\
10.4394390889339	0.143422413793103\\
12.9394390889339	0.0779966969446738\\
15.4394390889339	0.0302270714737508\\
17.9394390889339	0.00624327598740005\\
20.4394390889339	0.00044536761975492\\
22.9394390889339	4.87249706583206e-06\\
24.1894390889339	2.18014838634955e-07\\
25.4394390889339	0\\
27.9394390889339	0\\
};

\addplot [smooth, solid, color=green, mark=triangle, mark options={solid, green,scale=1.6}]
  table[row sep=crcr]{%
0.914246081470312	0.4583\\
3.41424608147031	0.434576923076923\\
5.91424608147031	0.388838235294118\\
8.41424608147031	0.322220338983051\\
10.9142460814703	0.245883620689655\\
13.4142460814703	0.161091659785301\\
15.9142460814703	0.0825433270082226\\
18.4142460814703	0.0291434456021323\\
20.9142460814703	0.00589588748607501\\
23.4142460814703	0.000495838816374435\\
25.9142460814703	8.37994536003107e-06\\
28.4142460814703	5.22032376447987e-09\\
30.9142460814703	0\\
};

\addplot [smooth, dashed ,color=black, mark=triangle, mark options={solid, black,scale=1.6}]
  table[row sep=crcr]{%
0.914246081470312	0.3947\\
3.41424608147031	0.352730769230769\\
5.91424608147031	0.304514705882353\\
8.41424608147031	0.242694915254237\\
10.9142460814703	0.182120689655172\\
13.4142460814703	0.123077621800165\\
15.9142460814703	0.0672507906388362\\
18.4142460814703	0.0256982069299733\\
20.9142460814703	0.0055694857036762\\
23.4142460814703	0.000488618984955721\\
25.9142460814703	8.37313239632372e-06\\
28.4142460814703	5.22032376447987e-09\\
30.9142460814703	0\\
};

\addplot [smooth,dashed, color=black, every mark/.append style={solid,fill=gray,draw=black,scale=1.2}, mark=*]
  table[row sep=crcr]{%
-4.36382637962824	0.457757757757758\\
0.636173620371758	0.372372372372372\\
5.63617362037176	0.259665070475881\\
10.6361736203718	0.144116338560783\\
15.6361736203718	0.0610754639531618\\
20.6361736203718	0.0149328432910522\\
25.6361736203718	0.00236143479386723\\
30.6361736203718	0.000113000113000113\\
};

\addplot [smooth,dashed , color=black, mark=o, mark options={solid,black,scale=1.2}]
  table[row sep=crcr]{%
-4.26392301147839	0.4375\\
0.73607698852161	0.345274509803922\\
5.73607698852161	0.172085820895522\\
10.7360769885216	0.0285298776097912\\
15.7360769885216	0.000551000555864369\\
20.7360769885216	8.04828973843058e-08\\
25.7360769885216	0\\
30.7360769885216	0\\
};

\addplot [smooth,dashed , color=black, mark=star, mark options={solid,scale=1.2}]
  table[row sep=crcr]{%
-4.26586046360554	0.4302\\
0.734139536394462	0.323647058823529\\
5.73413953639446	0.141\\
10.7341395363945	0.0198999280057595\\
15.7341395363945	0.000447887715397443\\
20.7341395363945	8.04828973843058e-08\\
25.7341395363945	0\\
30.7341395363945	0\\
};

\addplot[smooth,color=white,solid,mark=square,mark options={solid, black}]
  table[row sep=crcr]{%
        1 2\\
};\label{P1}

\addplot[smooth,color=white,solid, mark=diamond, mark options={solid, black,scale=1.2}]
  table[row sep=crcr]{%
        1 2\\
};\label{P2}
\addplot[smooth,color=white,solid,mark=triangle, mark options={solid, black,scale=1.2}]
  table[row sep=crcr]{%
        1 2\\
};\label{P3}

\addplot[smooth, color=white, mark=*, mark options={solid, fill=gray,draw=black, scale=1.2}]
  table[row sep=crcr]{%
        1 2\\
};\label{P4}

\addplot[smooth, color=white, mark=o, mark options={solid, black, scale=1.2}]
  table[row sep=crcr]{%
        1 2\\
};\label{P5}

\addplot[smooth, color=white, mark=star, mark options={solid, black, scale=1.2}]
  table[row sep=crcr]{%
        1 2\\
};\label{P6}

\node [draw,fill=white, font=\tiny, anchor= south west, at={(0,1e-06)}]{
\shortstack[l]{
\ref{P4}~{8-CPFSK, $M=5$ \cite{Bender_SPAWC2019}}\\
\ref{P3}~{Proposed FTN-CPM, $T_{\textrm{cpm}}=2T_{\textrm{s}}$}\\
\ref{P2}~{Proposed FTN-CPM, $T_{\textrm{cpm}}=1.6T_{\textrm{s}}$}\\
\ref{P1}~{Proposed FTN-CPM, $T_{\textrm{cpm}}=T_{\textrm{s}}$}\\
\ref{P5}~{4-CPFSK, $M=2$ \cite{Landau_CPM_2018}}\\
\ref{P6}~{4-CPFSK, $M=4$
\cite{Landau_CPM_2018}}}};

\end{axis}
\end{tikzpicture}%
\caption{BER performance of the considered CPM waveforms}
\label{fig:ber_cpm}
\end{center}
\vspace{-1em}
\end{figure}
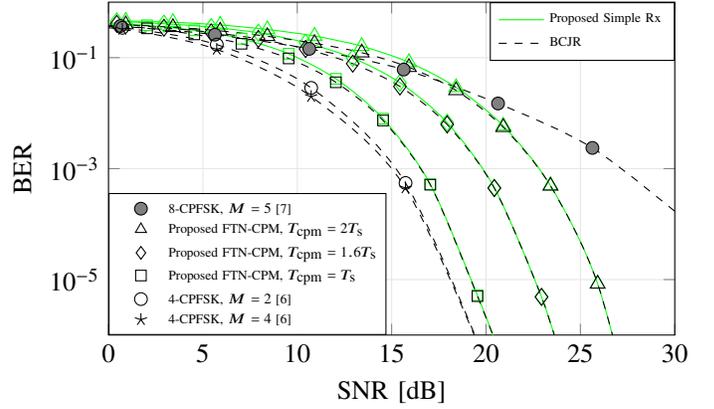

The uncoded BER is shown in Fig.~\ref{fig:ber_cpm}.
The increase of the length of the frequency pulse $T_{\textrm{cpm}}$ in the proposed binary CPM reduces the distance between the constellations points, which results in increased sensitivity to noise.
Different to the 1-bit customized 8-CPFSK \cite{Bender_SPAWC2019}, the proposed FTN-CPM shows a BER performance which decreases fast for higher $\mathrm{SNR}$.
An additional highlight is that the proposed simple receiver strategy  results in a BER performance which is almost identical with the performance of the optimal BCJR-based CPM demodulator especial at medium and high $\mathrm{SNR}$.
\section{Conclusions}


The present study introduces
a novel subclass of CPM signals, namely CPM signals with \textit{faster-than-Nyquist} signaling.
The novel waveform is especially promising in the context of 1-bit quantization at the receiver, because is privides a good steering of zero-crossings. 
By considering Carson's bandwidth criterion it is shown that a waveform equivalent to common CPFSK in bandwidth and information bits
can be constructed with a
higher signaling rate.
Numerical results show superior performance in terms of spectral efficiency and BER for the channel with 1-bit quantization at the receiver.
The illustrated binary FTN-CPM signal can be detected with an extremely simple detector with a performance close to MAP detection.

\bibliographystyle{IEEEtran}
\bibliography{bib-refs}

\end{document}